\documentclass[a4paper,fleqn,usenatbib,useAMS]{mnras}

\usepackage{graphicx}	% Including figure files
\usepackage{amsmath}	% Advanced maths commands
\usepackage{amssymb}	% Extra maths symbols
\usepackage{multicol}        % Multi-column entries in tables
\usepackage{bm}		% Bold maths symbols, including upright Greek
\usepackage{pdflscape}	% Landscape pages
\usepackage{epsf}
\usepackage{tikz}
\usepackage{hyperref}
\usetikzlibrary{decorations.pathreplacing, shapes, arrows}
\usetikzlibrary{fadings}

\newcommand{\kms}{\;{\rm km}\,{\rm s}^{-1}}
\newcommand{\kmsmpc}{\kms\;{\rm Mpc}^{-1}}

\newcommand{\HI}{{\sc Hi}}

\newcommand{\msolar}{\;{\rm M}_{\odot}}

\newcommand{\muf}{{\sc Mufasa}}

\newcommand{\blue}{\color{blue}}

\newcommand{\radpred}{\hyperref[rad18]{\sc{rad18}}}

\usepackage[T1]{fontenc}
\usepackage{ae,aecompl}

\usepackage{newtxtext,newtxmath}
\title[\HI~ in galaxies]{Classifying galaxies according to their \HI~content}

\author[Andrianomena, Rafieferantsoa  \& Dav\'e]{Sambatra Andrianomena$^{1,2}$
\thanks{e-mail: \href{mailto:andrianomena@gmail.com}
{andrianomena@gmail.com}},
Mika Rafieferantsoa$^{2,3,4}$
\thanks{e-mail: \href{mailto:rafieferantsoamika@gmail.com}
{rafieferantsoamika@gmail.com}},
Romeel Dav\'e$^{2,3,5}$
\\
\\$^1$ South African Radio Astronomy Observatory (SARAO), Black River Park, Observatory, Cape Town, 7925, South Africa 
\\$^2$ University of the Western Cape, Bellville, Cape Town 7535,
South Africa
\\$^3$ South African Astronomical Observatory, Observatory,
Cape Town 7925, South Africa
\\$^4$ Max-Planck-Instit\"ut f\"ur  Astrophysik, Garching, Germany
\\$^5$ Institute for Astronomy, Royal Observatory, Edinburgh EH9 3HJ, UK
}

\date{Last updated 2018 February 14; in original form 2018 February 13}

\pubyear{2018}

\begin{document}
\label{firstpage}
\pagerange{\pageref{firstpage}--\pageref{lastpage}}
\maketitle

\begin{abstract}
We use machine learning to classify galaxies according to their \HI~content, based on both their optical photometry and environmental properties. The data used for our analyses are the outputs in the range $z = 0-1$ from \muf\
cosmological hydrodynamic simulation. In our previous paper, where we predicted the galaxy \HI~content using the same input features, \HI~rich galaxies were only selected for the training. In order for the predictions on real observation data to be more accurate, the classifiers built in this study will first establish if a galaxy is \HI~rich ($\rm{log(M_{\HI}/M_{*})} > -2 $) before estimating its neutral hydrogen content using the regressors developed in the first paper. We resort to various machine learning algorithms and assess their performance with some metrics such as \textit{accuracy}, $f_{1}$, ROC AUC, \textit{precision}, \textit{specificity} and \textit{log loss}. The performance of the classifiers, as opposed to that of the regressors in previous paper, gets better with increasing redshift and reaches their peak performance around $z = 1$ then starts to decline at even higher $z$. Random Forest method, the most robust among the classifiers when considering only the mock data for both training and test in this study, reaches an accuracy above $98.6 \%$ at $z = 0$ and above $99.0 \%$ at $z = 1$, which translates to a ROC AUC above $99.88\%$ at low redshift and above $99.96\%$ at higher one. We test our algorithms, trained with simulation data, on classification of the galaxies in RESOLVE, ALFALFA and GASS surveys. Interestingly, SVM algorithm, the best classifier for the tests, achieves a \textit{precision}, the relevant metric for the tests, above $87.60\%$ and a \textit{specificity} above $71.4\%$ with all the tests, indicating that the classifier is capable of learning from the simulated data to classify \HI~rich/\HI~poor galaxies from the real observation data. With the advent of large \HI~21 cm surveys such as the SKA, this set of classifiers, together with the regressors developed in the first paper, will be part of a pipeline, a very useful tool, which is aimed at predicting \HI~content of galaxies. 
\end{abstract}


%%%%%%%%%%%%%%%%%%%%%%%%%%%%%%%%%%%%%%%%%%%%%%%%%%

%%%%%%%%%%%%%%%%% BODY OF PAPER %%%%%%%%%%%%%%%%%%

\section{Introduction}
\label{intro}
\indent

Much effort has been put into understanding the role of neutral hydrogen
in galaxy formation and evolution. In the canonical picture based on the
Hubble Sequence, the spiral galaxies are rich in cold gas and star forming,
whereas the ellipticals are red and quiescent.  However, an increasing number
of observational evidence shows that these correlations are not always true.
Local early-type galaxies from the ATLAS$^\mathrm{3D}$ survey were shown to
contain significant cold gaseous components \citep{Davis-11}. They found that
the relative angles between the gaseous and stellar planes show a bimodal
distribution, but found no plausible explanation for such difference.
%{\bf WHAT IS THE DICHOTOMY?  WHY IS THIS IMPORTANT/RELEVANT HERE?}.
This indicates that the gas distribution of a galaxy does not necessarily
follow that of the stellar component. Therefore, direct inference of the gas
content of galaxy based on its optical content is inaccurate.
Elliptical galaxies are observed to form stars in cool core massive clusters
\citep{Donahue-11} that is suggestive of the presence of cold gas in those
objects. The amount of gas components in massive ellipticals is crucial to
understanding the evolution and growth of galaxies at the massive end, but
the presence of kinematic abnormalities in their gas content as well as the
uncertain effects of the Active Galactic Nuclei (AGN) feedback can affect
the surface density of the gas content to pull the galaxies below the \HI\
detection limit, especially at higher redshifts.

Spiral galaxies are gas rich, but the limitations of observing the neutral
gas at intermediate redshift prevent a robust study of the evolution of their 
gas content.  Low redshift ($z\lesssim0.4$) \HI\ can be observed with the
21cm emission line to provide the neutral hydrogen mass distribution of
nearby galaxies. For instance, the Arecibo Legacy Fast ALFA \citep[ALFALFA;]
[]{Haynes-18} observed $\sim30000$ galaxy \HI\ fluxes.
The highest redshift galaxy ($z=0.376$) detected in 21 cm emission was
observed with the COSMOS \HI\ Large Extragalactic Survey (CHILES)
\citep{Fernandez-16}. At any substantially higher redshift, the \HI\ content
of galaxies is inferred from Damped Lyman Alpha systems (DLAs) in the spectra of
background quasars, but it is difficult to measure the \HI\ mass from DLAs, and the relationship between galaxies and DLAs is not completely clear.
The upcoming blind surveys such as Looking At the Distant Universe with the
MeerKAT Array (LADUMA)
%{\bf Spell out every acronym the first time you use it!}
on MeerKAT and eventually follow-up surveys on the SKA aim to measure the \HI\ content
of galaxies at intermediate redshifts, to $z\sim 1$ and beyond.

The gas content of satellite galaxies are substantially impacted by  environmental effects. 
Observationally, only $25\%$ of $\alpha.40$ \citep[ALFALFA 40\%;][]{Haynes-11}
galaxies were found to be in groups or clusters \citep{Hess-Wilcots-13}, which is lower than for the overall galaxy population.
They found that in contrast to increasing optical sources towards to the
center of groups or clusters, the number of \HI\ sources decreases.
This is also supported from theoretical views. Using hydrodynamical
simulation, \citet{Rafieferantsoa-15} showed that the fraction of \HI\ 
deficient galaxies increases towards higher halo masses. This is related to
the star formation quenching timescale decrease towards higher halo mass:
from $>3$ Gyr for $\mathrm{M_{halo}} <10^{12}\msolar$ to $<1$ Gyr for
$\mathrm{M_{halo}} > 10^{13}\msolar$~\citep{Rafieferantsoa-19}.
Recent observational work by \citet{Foltz-18} agrees with this prediction,
but in contrast \citet{Fossati-17} argues for no relationship between galaxy
quenching timescales and halo mass.  Simulations also suggest that the 
presence of \HI\
is strongly correlated with star formation, even if the star formation is physically occurring in molecular gas~\citep{Dave-17a}.
Therefore, the \HI\ content appears to have a complex relationship with
respect to stellar mass, star formation rate, morphology, and environment.  This makes it challenging to predict what the \HI\ content of any given galaxy will be without accounting for the full range of its properties.

In order to better design and interpret upcoming \HI\ surveys, it is useful to be able to estimate the expected \HI\ content of galaxies that will be observed based on their already-measured multi-wavelength properties.
To do so, here we develop and employ galaxy classification tools using machine learning.
Galaxy classification is a very useful approach as it can provide insights
into the physical processes by which galaxies evolve over cosmic time.
There exist different and complementary ways to classify galaxies depending
on the availability of the data, for instance morphological classification
or spectral classification.  The Hubble Sequence focuses on morphological
classification, while spectral classification via absorption and emission
lines provides more information about the chemical composition and stellar
populations of galaxies \citep{morgan1957spectral}.
\cite{zaritsky1995spectral} developed a $\chi^2$-fitting approach
to identify the best linear combination of template spectra that matches
the observed spectrum in order to classify galaxies spectroscopically with low
signal to noise ratio (S/N), and found good correlations of $\geq 80\%$
between spectra and morphology from Hubble classification.
\cite{slonim2001objective} presented a novel \textit{information bottleneck}
(IB) approach, improving on the then-standard geometrical and statistical
approaches, to classify galaxy spectra using 2dF Galaxy Survey
\citep{colless1998looking, folkes19992df}.
In a seminal work, \cite{fukugita2007catalog} conducted morphological
classification of galaxies which was achieved by simple visual inspection
where volunteers catalogued thousands of objects from Sloan Digital Sky
Survey Data Release 3 \citep[SDSS DR3;][]{york2000sloan}
in order to obtain the rate of interacting galaxies. The need for automated
classification arose with the increasing amount of available survey data,
and it was demonstrated by \cite{naim1995automated} and
\cite{lahav1996neural} that accuracy achieved by a trained
\textit{Artificial Neural Network} in classifying galaxies is comparable
to that of a human expert. In a morphological classification of high redshift
galaxies that \cite{HuertasCompany:2007xa} conducted using
\textit{Support Vector Machines}, they argued that at $z > 1$ early type
galaxies were underestimated in the classifications using sample from COSMOS
HST/ACS \citep{koekemoer2007cosmos} owing to the effects of morphological
\textit{k}-correction. In galaxy morphological classification, tree-based algorithms have
also proved to be relatively robust classifier compared to other machine
learning algorithms, as reported by \cite{gauci2010machine}.  Hence there is a long history of using sophisticated galaxy classification methods in astronomy, but so far this has not been extensively applied to studying \HI.

In our previous work in {\citet{rafieferantsoa2018predicting}\label{rad18}}
(\radpred\ hereafter), we investigated the possibility of estimating the
\HI~content of galaxies using a variety of machine learning algorithms.
Considering both the optical and environmental properties of the galaxies as
input features, the algorithms were trained using large subsets of data from
\muf\ simulation and tested on different subsets. They found that the
performance of all regressors -- assessed by using root mean squared error
({\sc{rmse}}) and Pearson's correlation coefficient (\textbf{r}) as metrics
-- degraded at higher redshift. Despite the tendency of all learners to
under-predict the high \HI~richness and over-predict the low one, random
forest method -- followed tightly by deep neural network --  exhibited an
overall best performance; achieving an {\sc{rmse}}~$\sim 0.25$ (corresponding
to $\textbf{r} \sim 0.9$) at $z = 0$. They then applied the regressors to
real data from two different surveys, RESOLVE and ALFALFA. To this end,
they trained the algorithms with an output from \muf\ at $z = 0$ and used
them to predict the \HI~content of galaxies from real observations.
Their results proved that the learners which they built can be potentially
used for \HI~study with the upcoming large \HI~surveys like the SKA.
Prior to this work, related study by \citet{Teimoorinia-17} also investigated
the estimation of \HI\ content of galaxies based on the SDSS and ALFALFA data
using 15 derived galaxy parameters.

However, in \radpred\ we only considered \HI~rich galaxies ($\rm{log(M_{\HI}/M_{*})} \geq -2 $), hence the machine learning methods were trained to predict the gas content of \HI~rich galaxies only.  Therefore, at this stage, those algorithms on their own can't be deployed in real world application where not all galaxies will be \HI~rich.  Models generally predict that galaxies are bimodal in their \HI\ content, particularly since satellite galaxies lose their \HI\ quite rapidly, after a delay period, once they enter another halo~\citep{Rafieferantsoa-19}.  To extend our work to be more generally applicable, we therefore need a way to classify galaxies as \HI~rich or \HI~poor based on available photometric data.

In this follow-up paper, we address this issue by building a set of learners that filter out the \HI~poor galaxies in real survey, such that the regressors built in \radpred\ only predict galaxy gas content known to be above a certain threshold. Together with the classifiers, the regressors will form a pipeline which will be used to estimate \HI~gas of galaxies in real observation. The approach is to use the same set of input features as in \radpred\ for the classification.  This paper thus extends our approach to be more generally applicable to any galaxy survey that contains the requisite input features, which are chosen to be typically observationally accessible in present and upcoming multi-wavelength surveys.

We present our machine learning setup for our analyses in \S\ref{setupmachine} and list all the algorithms we consider in \S\ref{algorithms}. The results are shown in \S\ref{resultsclass} and we demonstrate how the methods can be applied to data from real surveys in \S\ref{applicationdata}. We finally conclude in \S\ref{conclusiondiscussion}.

%-------------------------------------------------------------------------------

\section{Setups}\label{setupmachine}
It is first noted that we make use of the same outputs ($z = 0 - 1$) from \muf\ simulation to build our classifiers. Considering the Planck cosmological parameters  $\Omega_m =
0.3$, $\Omega_\Lambda = 0.7$, $\Omega_b = 0.048$, $H_0 = 68 \kmsmpc$, $\sigma_8 = 0.82$ and $n_s = 0.97$ \citep{-16}, each snapshot results from simulating a comoving box of 50$h^{-1}$Mpc with a resolution of N = $512^3$ for each species (dark matter and gas). For the training, the features $\{u,g,r,i,z,U,V,J,H,K_s, \Sigma_3, v_{gal}\}$ are considered whereas our target -- as in the case of a binary classification -- is one of the two classes;  \textbf{0} to denote \HI~depleted galaxies ($\rm{log(M_{\HI}/M_{*})} < v_{thresh} $) and \textbf{1} for \HI~rich galaxies ($\rm{log(M_{\HI}/M_{*})} \geq v_{thresh} $). To split the galaxies into two classes, one simply needs to run through all galaxies in the data and assign \textbf{0} or \textbf{1} to it if its gas content is below or above the threshold value v$_\mathrm{thresh}$. In our case, we adopt v$_\mathrm{thresh} = -2$, {\it i.e.} the \HI\ content is 2 orders of magnitude fewer than the stellar content.

\begin{table*}
 \caption{List of all the setups that are considered in the analysis.
 For easy reference, each setup has been given a name.}
 \label{setups}
 \begin{tabular}{lllll}
  \hline
  Name & Surveys & Features & Target & Description\\
  \hline
  \hline
  fSMg  & SDSS & $u,\> g,\> r,\> i,\> z,\> v_{gal},\> \Sigma_{3}$
            & $\rm log(M_{HI}/M_{*})$ & redshift information not required\\[2pt]
  fSClr  & SDSS & \verb'color indices',$\>v_{gal},\> \Sigma_{3}$
            & $\rm log(M_{HI}/M_{*})$ & redshift information not required\\[2pt]
   fSCmb & SDSS
               & \verb'color indices', $u,\> g,\> r,\> i,\> z,\> v_{gal},\> \Sigma_{3}$
               & $\rm log(M_{HI}/M_{*})$ & redshift information not required\\[2pt]
  fAMg    & SDSS+Johnson+2MASS
              & $H,\>J,\> Ks,\> U,\> V,\>u,\> g,\> r,\> i,\> z,\>v_{gal},\> \Sigma_{3}$
               & $\rm log(M_{HI}/M_{*})$ & redshift information not required \\[2pt]
  fAClr  & SDSS+Johnson+2MASS
            & \verb'color indices',$\>v_{gal},\> \Sigma_{3}$ & $\rm log(M_{HI}/M_{*})$    
            & redshift information not required \\[2pt]
  \hline
  zSMg & SDSS & $u,\> g,\> r,\> i,\> z,\> v_{gal},\> \Sigma_{3}$
            & $\rm log(M_{HI}/M_{*})$ & prediction at a given redshift bin\\[2pt]
  zSClr & SDSS & \verb'color indices',$\>v_{gal},\> \Sigma_{3}$
            & $\rm log(M_{HI}/M_{*})$ & prediction at a given redshift bin\\[2pt]
  zSCmb & SDSS
               & \verb'color indices', $u,\> g,\> r,\> i,\> z,\> v_{gal},\> \Sigma_{3}$
               & $\rm log(M_{HI}/M_{*})$ & prediction at a given redshift bin \\[2pt]
  zAMg & SDSS+Johnson+2MASS
            &$H,\>J,\> Ks,\> U,\> V,\>u,\> r,\> r,\> i,\> z,\>v_{gal},\> \Sigma_{3}$
            & $\rm log(M_{HI}/M_{*})$ &prediction at a given redshift bin \\[2pt]
  zAClr & SDSS+Johnson+2MASS
            & \verb'color indices',$\>v_{gal},\> \Sigma_{3}$
            & $\rm log(M_{HI}/M_{*})$ & prediction at a given redshift bin \\[2pt]

  \hline
 \end{tabular}
\end{table*}

As in \radpred, we adopt different setups both in terms of features and type of training which we present again in Table~\ref{setups} for reference. For ``$z-$training'', a classifier is built at each redshift bin whereas for ``$f-$training'' we make use of all data available in the range $z = 0 - 0.5$. In contrast with the $f-$training in \radpred, we do not go to higher $z$ to train the learner. In all cases, $75\%$ of the data is used for training and the remaining is used for testing.

\section{Algorithms}\label{algorithms}
We used a rather wide variety of machine learning algorithms in \radpred\ to see which one captures best the features from the data in order to make good predictions. Having gained a better understanding about how the methods dealt with information from the data, we consider most of them for this classification problem. It is worth reiterating that as opposed to regression task where the label is a numerical variable, the label for a classification task is a class -- represented by integers mainly\footnote{Categorical variable.}. 

\textit{k-Nearest Neighbour (kNN) - Classification}: the principle remains the same as in regression but instead of averaging the targets of $k-$closest neighbours  to make prediction, the predicted class $y_{new}$ of a new instance $\textbf{x}_{new}$ is simply the majority of the classes of $k-$neighbours of $\textbf{x}_{new}$.

\textit{Random forest (RF) and Gradient boosting (GRAD) - Classification:} decision tree is still the base estimator of both RF and GRAD. In contrast with its regressor counterpart, the decision tree classifier splits the training set at a split point $s_{i}$ using a feature $i$. The splitting is done in such a way as to minimize the objective function
\begin{equation}
\mathcal{F} = \frac{n_{R_{1}}}{n}G_{R_{1}} + \frac{n_{R_{2}}}{n}G_{R_{2}}, 
\end{equation}
where $n_{R_{1}}$ is the number of examples in region $R_{1}$ and $n_{R_{2}}$ the number of examples in $R_{2}$. The total number of instances $n$ before the split is simply $n = n_{R_{1}} + n_{R_{2}}$. The Gini impurity\footnote{Also called Gini index.} $G$ of each region is given by 
\begin{equation}
G = 1 - \sum_{i = 1}^{k}p^{2}_{i},
\end{equation}
where $p_{i}$ is the probability of an instance to belong to a class $i$ in the region. This can be computed by the ratio between  the number of intances belonging to a class $i$ and the number of all instances in the region. The splitting can be done recursively on the resulting nodes depending on the required size of the tree. The RF method predicts the class of a new instance $\textbf{x}_{new}$ by aggregating the predictions of all its decision trees. The expression of the GRAD classifier is quite similar to Eq.~{\blue{6}} in \radpred.

\textit{Deep neural network (DNN) - Classification:} In contrast with the DNN regressor, the activation function of the output layer is a sigmoid function\footnote{Also named \textit{logit}.} 
\begin{equation}
\sigma(x) = \frac{1}{1 + e^{-x}},
\end{equation} 
which computes the probabibility $p_{i}$ that an instance belongs to class $i$. In this case specifically, if $p \geq 0.5$, $y_{new}$ is $\textbf{1}$ (positive class) whereas for $p < 0.5$ $y_{new}$ is $\textbf{0}$ (negative class). The objective function, known as \textit{log loss}, is defined as
\begin{equation}\label{logloss}
\mathcal{F} = -\frac{1}{N}\sum_{i=1}^{N}y_{i}{\rm{log}}(p_{i})+(1 - y_{i}){\rm{log}}(1-p_{i}).
\end{equation}
The weights and biases are updated via backpropagation as usual. The cost function in Eq.~\ref{logloss} can be generalised for multiclass case by using what is called \textit{cross entropy} defined as
\begin{equation}
\mathcal{F} = -\frac{1}{N}\sum_{i=1}^{N}\sum_{k=1}^{\mathcal{K}}y_{k}^{i}{\rm{log}}(p_{k}^{i}),
\end{equation}
where $\mathcal{K}$ is number of classes.
\section{Galaxy classification}\label{resultsclass}
The objective in this work is to be able to establish whether a galaxy is \HI~rich or \HI~poor by exploiting both its optical and environmental data. To do so, we build various classifiers (see~\S\ref{algorithms}) and compare their performance qualitatively using various metrics which we present now along with some useful terminology in machine learning.

\begin{figure}
\includegraphics
[width=0.9\columnwidth]
{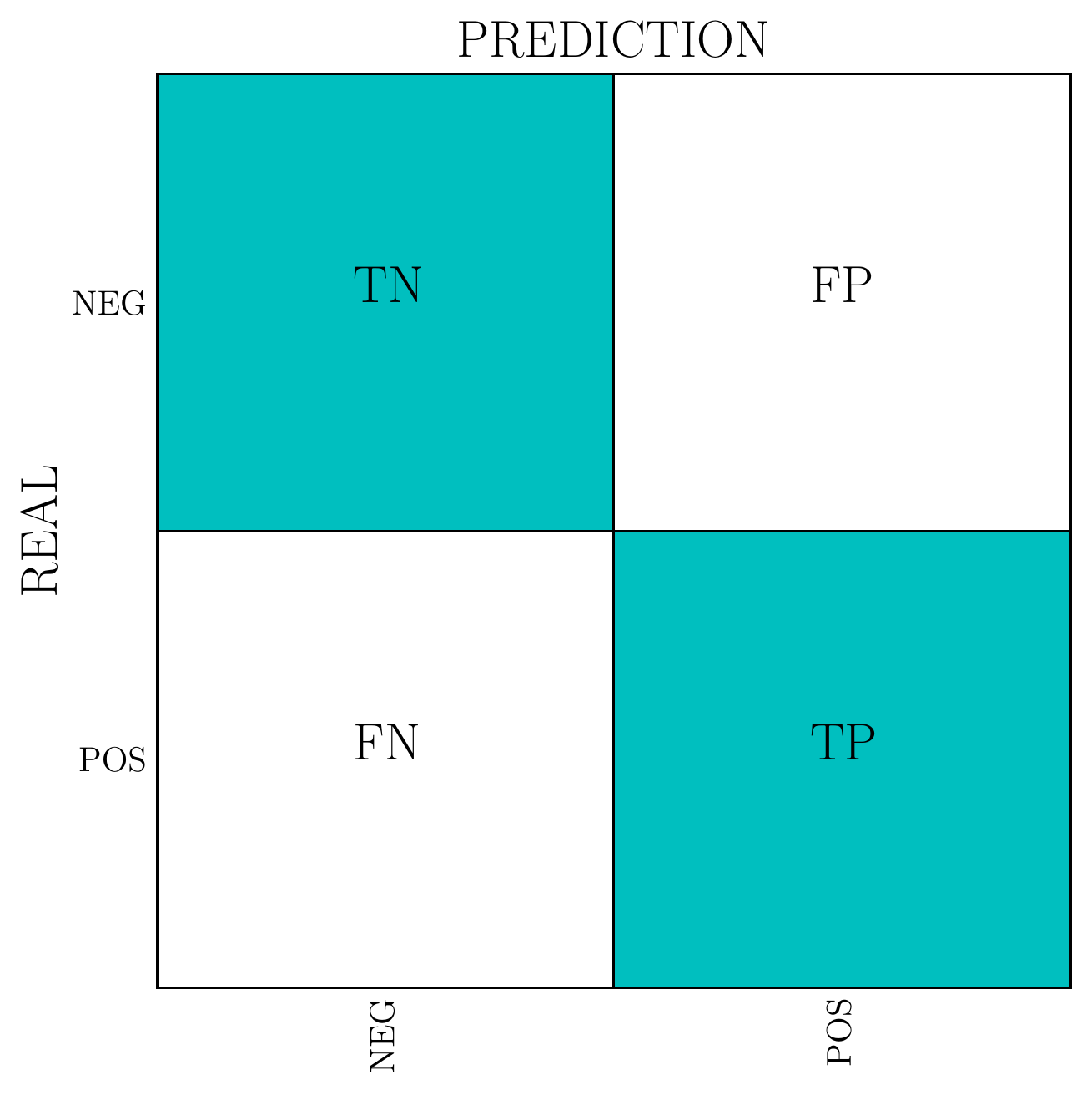}
\caption{Confusion matrix $2 \times 2$ for a binary classification. \textit{Negative class} is \HI~poor, \textit{Positive class} is \HI~rich.}
\label{confusionmatrix}
\end{figure}

\textit{Accuracy:} In binary classification\footnote{And even in mutliclass case.}, it measures the ratio of the correct predictions on a test sample, \textit{i.e.}
\begin{equation}
accuracy = \frac{TP + TN}{FN + FP + TP + TN},
\end{equation}
where $TP$ and $TN$ are \textit{True Positive} -- number of instances that are correctly predicted by the classifier to belong to \textbf{1} -- and \textit{True negative} -- number of instances that are correctly predicted by the classifier to belong to \textbf{0} -- respectively. {\it FN} or \textit{False Negative} denotes the number of instances that belong to \textbf{1} but are classified as \textbf{0} and {\it FP} or \textit{False Positive} indicates the number of instances that belong to \textbf{0} but are predicted as \textbf{1}. A confusion matrix, which is represented in Figure.~\ref{confusionmatrix}, is $2 \times 2$\footnote{$n \times n$ in multiclass case.} matrix which summarizes the predictions of a classifier on a test set. \\

\textit{Precision:} It indicates how well the algorithm minimizes the number of instances incorrectly identified as a \textit{Positive class} ({\it FP}) and is given by
\begin{equation}\label{precision}
precision = \frac{TP}{TP + FP}.
\end{equation}
A good precision (high value close to one) translates to low {\it FP}.\\

\textit{Recall:} Also called \textit{sensitivity}, it characterizes the ability of the method to minimize the number of instances wrongly identified as a \textit{Negative class} ($FN$). It is given by
\begin{equation}\label{recall}
recall = \frac{TP}{TP + FN}.
\end{equation}
It is worth noting that, provided a classifier, if {\it FP} increases then {\it FN} decreases and vice versa. In other words, an increase in \textit{precision} implies a decrease in \textit{recall} -- the so called \textit{precision-recall} tradeoff. In our case, since we are mainly interested in identifying \HI~rich galaxies whose gas content is to be predicted by our regressors built in \radpred, we require our classifier to have good \textit{precision}, as having a learner with a lower {\it FP} (hence higher {it FN}) -- lower number of \HI~poor galaxies predicted to belong to class of \HI~rich galaxies -- is in our case more preferable than a learner with a lower {\it FN}, hence higher {\it FP}.\\

\textit{$F_{1}$ score:} This metric which combines $precision$ and $recall$ is their harmonic mean, given by
\begin{equation}
F_{1} = \frac{TP}{TP + \frac{FN + FP}{2}}.
\end{equation} 
High $F_{1}$ score simply means that both $precision$ and $recall$ are also high, which is the ideal case.\\

\textit{Log Loss:} This quantity, given by Eq.~\ref{logloss}, is also used as a metric. The lower its value, the better the classifier is. \\
 
\textit{Receiving Operating Characteristic - Area Under the Curve (ROC AUC):} It is also possible to plot \textit{recall} against \textit{FP} rate which is given by $1 - specificity$ where $$specificity = \frac{TN}{TN + FP}.$$As can be seen from Eqs.\ref{precision}-\ref{recall}, $FP$ follows the increase of \textit{recall} as a consequence of the \textit{precision}-\textit{recall} trade-off. Another measure of the performance of a classifier is then to compute the area under the curve (\textit{recall} \textit{vs} \textit{FP} rate). A perfect learner would have \textit{ROC AUC} = 1. 

A binary classifier uses a threshold parameter such that a new instance will be classified as \textit{positive} or \textit{negative} if the predicted probability is above or below the threshold respectively. A \textit{precision-recall} (alternatively \textit{recall-FP} rate) pair corresponds to a single value of a threshold parameter of a classifier and the idea behind the \textit{ROC} curve is to find the best pair values \textit{precision-recall} (alternatively \textit{recall-FP} rate) in order to mitigate the trade-off between them, \textit{i.e.} finding a threshold parameter value of the classifier such that both \textit{precision} and \textit{recall} are high.  
The results are now presented in the following.

\subsection{Dependence on redshift}

Table~\ref{setups} lists the various setups that we feed to our machine learning algorithms.  The name specifies whether it is  uses $f-$training or $z-$training, whether we use SDSS data only (S) or all data including near-IR (A), and whether we use magnitudes (Mg) or colors (Clr) or combine them (Cmb).  In all cases we use environment as measured by the third nearest neighbor ($\Sigma_3$), as well as the galaxy peculiar velocity ($v_{gal}$).

\begin{figure*}
%Correlation_plot_all.py
\includegraphics{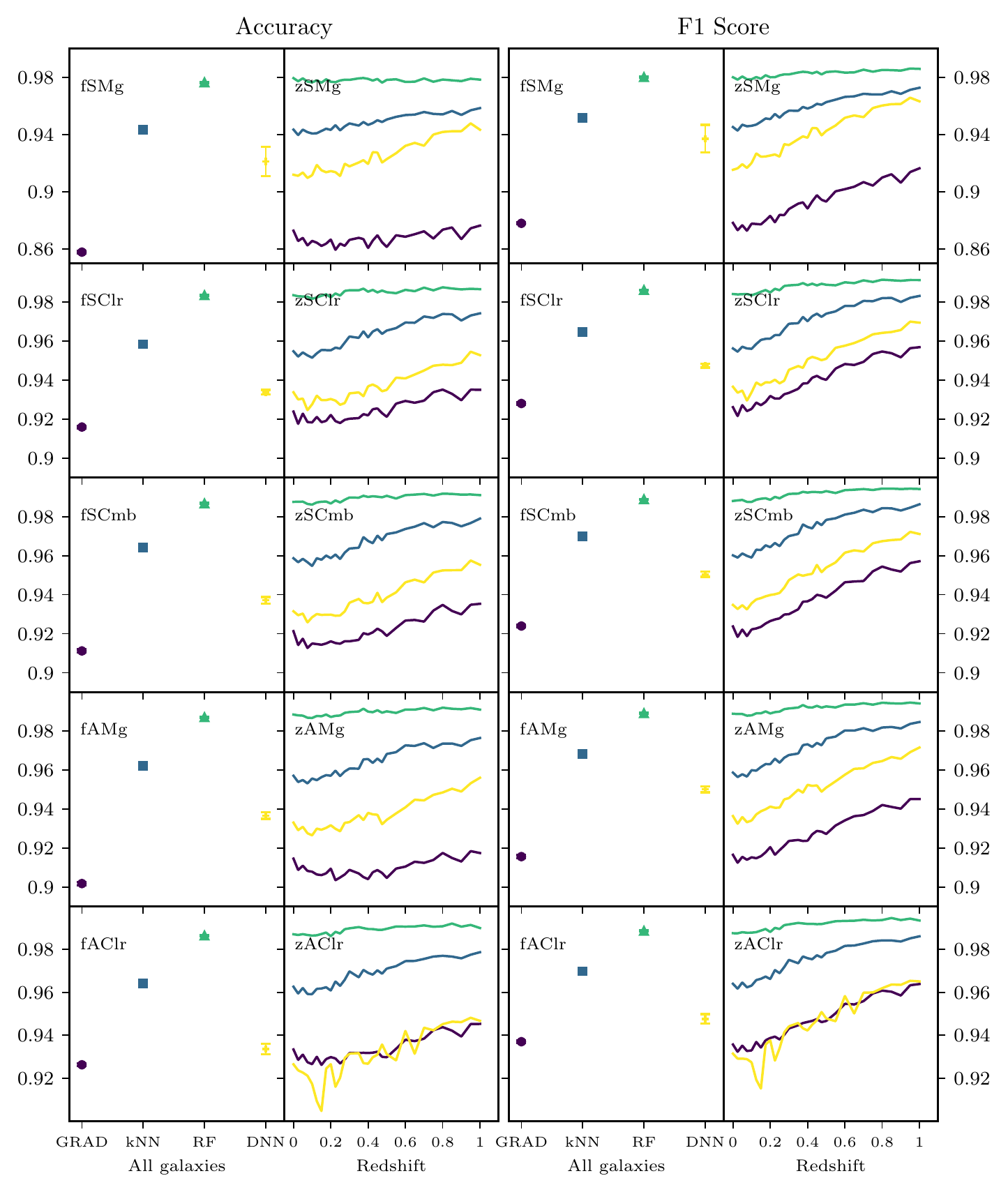}
\caption{\textit{Accuracy} and $f_{1}$ are shown on the 2 columns from the left and right,
respectively. Good performance means high values of  both \textit{accuracy} and $f_{1}$. The dots, color coded
by the training models we use, represent the performance (\textit{accuracy} and $f_{1}$) of each classifier trained on all the data available between $z=0-0.5$, ``$f-$training''. In the same way, the lines denote the value of the two metrics of each learner as a function of redshift ``$z-$training''. Each row shows different results for different
setups. The \textit{accuracy} values are shown on the left y-axes and the $f_{1}$ values
on the right y-axes.}
\label{fig_correlationall}
\end{figure*}

In Figure.~\ref{fig_correlationall}, we show the results corresponding to each classifier selected in our investigation, considering only two metrics here, \textit{accuracy} and $f_{1}$, for illustration purpose. The first column shows the accuracy achieved by each method with different input features for ``$f-$training'', the second column is the resulting accuracy for ``$z-$training'', the third column presents the $f_{1}$ score for ``$f-$training'' and finally the fourth one is the $f_{1}$ score for ``$z-$training''. 

Most classifiers attain accuracy and $f_{1}$ scores exceeding 0.9, which indicates that it is robustly possible to classify galaxies into \HI~rich vs. \HI~poor based on observable properties, at least in the idealised case of training and testing on simulated data alone.
Still, there are clear differences among the classifiers.
Random forests (RF; green) clearly exhibits the best performance whereas GRAD (purple) is relatively the weakest. For instance, RF (``$z-$training'') $accuracy$ and $f_{1}$ both reach $\sim 0.98$ at $z = 0$ and $\sim 0.99$ at $z = 1$, with similar values when combining data from $z=0-0.5$ (``$f-$training").  kNN shows values $\sim 0.95$, while DNN's performance is consistently poorer.

The dependences of both \textit{accuracy} and $f_{1}$ on redshift follow similar trend; they both increase as we go to higher $z$. This indeed looks very promising, since the improving performance of the classifier with increasing $z$ may compensate for the decreasing performance of the regressor built in \radpred\ at higher $z$, although this is only valid up to $z \sim 1$ since the performance of the classifier reaches their peak around that redshift then starts to degrade. That limitation is the reason we only show the results up to $z\sim1$. In other words, most of the \HI~poor galaxies can be filtered out by the classifier such that the regressor will only estimate the gas content of the \HI~rich galaxies. 

As expected, the value of the \textit{accuracy} and that of $f_{1}$ when training the learners with all the data available between $z = 0-0.5$ is approximately the average of \textit{accuracy}'s and that of $f_{1}$'s within that $z$-bin. As already mentioned in \radpred\, the main idea behind the ``$f-$training'' is to anticipate the fact that in real observations, retrieving redshift information is not an easy task. Therefore  we make an attempt at also building a classifier without relying on redshift information. The high values of both \textit{accuracy} and $f_{1}$ $\sim 0.9$ for all learners with any setup except fSMg demonstrate that it is indeed possible to build a relatively good classifier without taking into account redshift information. 

\subsection{Dependence on input features}

We now look in more detail at how the classification is affected by the selected input features, i.e. comparing the rows in Figure.~\ref{fig_correlationall}.
In realistic scenarios, it is not always possible to have all the features available. This leads us to investigate different scenarios by considering different combinations of features. The best classifier (RF) does appear to be insensitive to the choice of input features with values of \textit{accuracy} and $f_{1}$ $\geq 0.98$ at all redshift bins, which is good news. However, for the learner with the worst performance (GRAD), it does not seem to be the case as its performance measures fluctuate with respect to the setup considered and are at their lowest values with zSMg setup (at $z = 0$, $f_{1}$ and \textit{accuracy} are both $\sim 0.87$; $z = 1$, $f_{1}$ $\sim 0.91$ and $accuracy \sim 0.88$) for ``$z-$training'' and $accuracy \sim 0.857$ and $f_{1}$ $\sim 0.877$ for ``$f-$training'' fSMg. 

\begin{figure}
\includegraphics
[width=\columnwidth]
{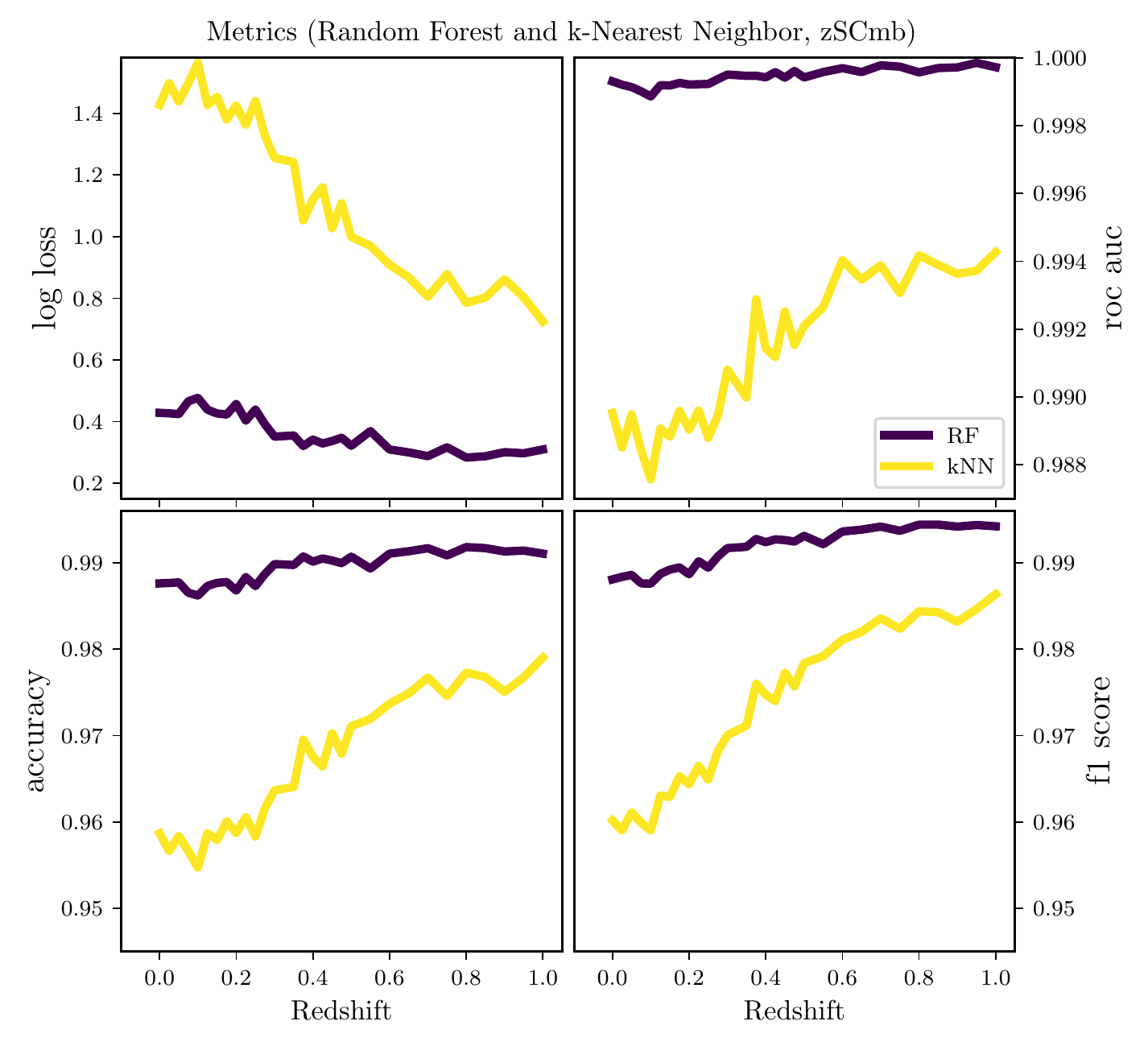}
\caption{Four metrics as a function of $z$ of both RF and kNN methods for zSCmb setup. \textit{Top left:} \textit{log loss}, \textit{top right:} \textit{ROC AUC}, \textit{bottom left: accuracy} and \textit{bottom right:} $f_{1}$.}
\label{fourmetricsrf}
\end{figure}

In Figure.~\ref{fourmetricsrf}, we show other metrics of the RF, namely \textit{ROC AUC} and \textit{log loss}, as function of redshift for zSCmb. As expected, the better performance at higher redshift bin corresponds to a lower \textit{log loss}. A \textit{ROC AUC} $\geq 0.99$ at all redshifts corroborates the fact that RF is our best classifier for this ideal scenario where classifiers are both trained and tested with mock data.

It is noted that the effects of the class imbalance -- potential issue owing to a big difference between the number of instances of each class in the training set which might cause a classifier to fail to label new instances in the test set properly -- have been checked by compensating the imbalance using {\sc{imblearn}}. No noticeable difference\footnote{If not the exact same results.} has been found between the two cases -- with and without compensation -- by comparing their resulting metrics. 
\subsection{Effects of setting up the classes}
In our main analyses, the \HI~galaxies are split into two distinct classes according to whether their \HI~gas masses are above or below a threshold of 0.01 times their stellar masses. The threshold value is broadly in accordance with observational \HI\ fraction limits. However, other classifications are possible.  Here we explore the impact of changing the classification metric.

\begin{figure}
%illustration.py
\includegraphics
[width=\columnwidth]
{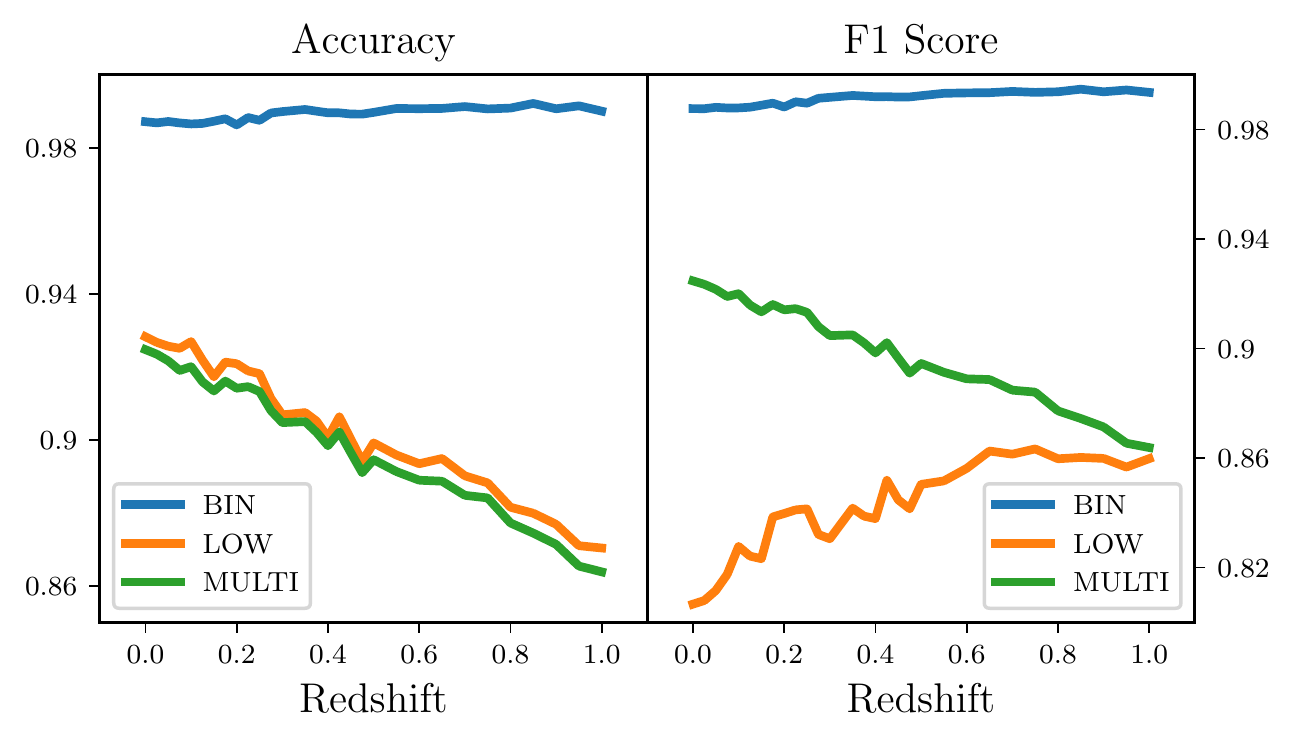}
\caption{\textit{Left panel: accuracy} as function of $z$, \textit{right panel:} $f_{1}$ as a function of $z$. Blue is for {\sc{BIN}}, orange for {\sc{LOW}} and green for {\sc{MULTI}}. The results are all related to RF algorithm.}
\label{twometricsclasses}
\end{figure}

We consider three new classification schemes.
\begin{itemize}
\item The Galex Arecibo SDSS Survey \citep[GASS;][]{Catinella-13} set a threshold limit of $\log(\mathrm{M_{HI}}) = 8.7$ for galaxies with $\mathrm{M_*}<10^{10.5}\msolar$ and $\log(\mathrm{M_{HI}/M_*}) = -1.8$ otherwise. However, in order to be consistent with the threshold value of gas fraction used in \radpred\ to denote \HI~depleted galaxies, we set it to be $\log(\mathrm{M_{HI}/M_*}) = -2$. We call this type of splitting BIN.
\item Another potential classification may be on whether a galaxy has higher \HI\ mass than stellar mass.  In this case, the classes are given by $\Big\{\rm{log(M_{\HI}/M_{*})} < 0 \rightarrow \textbf{0};\>\rm{log(M_{\HI}/M_{*})} \geq 0 \rightarrow \textbf{1} \Big\}$. We name this type of splitting LOW. 
\item Finally, we attempt splitting into {\it three} classes, as follows: $\Big\{\rm{log(M_{\HI}/M_{*})} < -2 \rightarrow \textbf{0};\> -2 \leq \rm{log(M_{\HI}/M_{*})} < 0 \rightarrow \textbf{1};\> \rm{log(M_{\HI}/M_{*})} \geq 0 \rightarrow \textbf{2} \Big\}$, which we call MULTI.
\end{itemize}
In Figure.~\ref{twometricsclasses}, we compare the results corresponding to the RF method when considering three types of splitting, namely BIN (blue), LOW (orange) and MULTI (green). For brevity we only consider RF, since it is our best classifier, and $z-$training since the $f-$training values are expected to be similar.

Overall, both \textit{accuracy} and $f_{1}$ are $\geq 0.80$ for all three types of splitting at all $z$ bins and it is quite clear that the algorithm performs best with our main type of splitting \HI~poor/\HI~rich, namely BIN.  It is also interesting to see that the \textit{accuracy} decreases with increasing redshift for both LOW and MULTI whereas $f_{1}$ increases as we go at higher redshift for LOW. Based on $accuracy$, the method performs similarly for LOW and MULTI splittings, but the difference in performance of the algorithm is striking when considering $f_{1}$ as a metric.  This indicates that the classifier performance does depend on the classes chosen, but for our purposes of separating \HI~rich and \HI~poor galaxies, it performs very well even with minor changes to the scheme\footnote{Slight change to the gas fraction limit.}.

It is worth noting that in this idealised case and in the light of the results in \radpred\, we did not include SVM method. However, as will be shown later, we include it for the different tests on real observation data.

\section{Application to observational data}\label{applicationdata}
The lack of available data is one of the drawbacks of using machine learning when solving a problem, be it regression or classification. To mitigate that issue, in the context of \HI~study, we aim at building classifiers trained with mock data from simulation and using them to identify \HI~rich galaxies in real surveys. 

As already demonstrated in \radpred\, the regressors that they built were able to learn from the mock data in order to predict the \HI~content of the galaxies from both RESOLVE and ALFALFA. Our approach here is to redo the same exercise but for a classification task, \textit{i.e.} training some classifiers with \muf\ data and utilising them to identify \HI~rich galaxies from the same surveys, RESOLVE and ALFALFA. In this study, we also consider another survey, GASS, in which both \HI~poor and \HI~rich galaxies are better represented for our tests. For the description of the first two surveys, we refer the interested reader to \radpred, and will now give a brief description of GASS.

\subsection{\textit{GALEX} Arecibo SDSS Survey (GASS) data}
GASS was aimed at investigating \HI~properties of a selected sample of galaxies ($\sim$ 1000) with available optical properties. The last data release (DR3) \citep{catinella2013galex}, which we use in our analyses, was built upon the first two data releases
\citep{catinella2010galex, catinella2012galex}. Within a relatively large volume survey of 200 Mpc already probed by SDSS primary spectroscopy survey, the \textit{GALEX} Medium Imaging Survey and ALFALFA, galaxies have stellar masses of  $10 < \rm{log}(M_{*}/\rm{M_{\odot}}) < 11.5$ that encompasses the transition mass.
The targets have \HI~richness above the detection limit of $0.015$ for $10.5 < \rm{log}(M_{*}/\rm{M_{\odot}})$ and a fix \HI~mass of $10^{8.7} \rm{M_{\odot}}$ for lower stellar masses.
The targets are designed to fall within $0.025<z<0.05$.

Using the Arecibo radio telescope, \cite{catinella2013galex} compiled a sample which has a fairly good representation\footnote{GASS \textit{representative} sample as they call it.} in which $62\%$ are referred to as \textit{detections} and the remaining $38\%$ as \textit{non detections}. The latter represent galaxies in which a relatively small gas mass fraction was observed hence required a longer integration time (but not more than 3h), whereas the former was found to have relatively large amount of gas mass fraction. For our analyses, we retrieved all the optical properties of each galaxy in the sample from SDSS database using their SDSS-ID. In order to have a more balanced test sample, we then split the sample into two classes: \HI~poor galaxies (class \textbf{0}) are those with $\rm{log}(M_{HI}/\rm{M_{*}}) < -1.55$ and the remaining are \HI~rich galaxies (class \textbf{0}). With this type of splitting, we have $56.8\%$ of the sample \HI~rich and the remaining \HI~poor.
\subsection{Testing the built classifiers}
We consider four different tests according to both the survey and input features
\begin{itemize}
\item \textbf{TEST 1}:~RESOLVE DATA, \verb'color indices' from all the band magnitudes available; SDSS (\textit{u,g,r,i,z}), 2MASS (\textit{J,H,K}), GALEX (\textit{NUV}) and
UKIDSS (\textit{Y,H,K})
\item \textbf{TEST 2}:~RESOLVE DATA, \verb'color indices' from only SDSS (\textit{u,g,r,i,z}) photometric data.
\item \textbf{TEST 3}:~ALFALFA, \verb'color indices' from only SDSS (\textit{u,g,r,i,z}) photometric data.
\item \textbf{TEST 4}: GASS data, \verb'color indices' from only SDSS (\textit{u,g,r,i,z}) photometric data.
\end{itemize} 
In all cases we split the simulated data for training and the considered test set into two categories, \HI~poor (class \textbf{0}) and \HI~rich (class \textbf{1}).
Our results are summarised in Table \ref{resolvealfalfa} and shown in Figure \ref{summaryresults}.

\subsection{TEST 1} 
The training set is composed of the data of snapshot at $z = 0$ from \muf\ since the galaxies to be classified in RESOLVE survey are all at present epoch. We make use of all the photometric data available in RESOLVE, \textit{i.e.} $\{u, g, r, i, z, J, H, Ks, NUV\}$.  We consider 5 metrics -- \textit{accuracy}, $f_{1}$, \textit{ROC AUC,
precision} and \textit{specificity}.  The results of the classification
from the learners selected in this work are presented in
Table~\ref{resolvealfalfa} and similarly shown in Figure
\ref{summaryresults}.

DNN has the highest \textit{accuracy} amongst
the algorithms followed by RF. This is reminiscent to the results found
in the regression problem in \radpred. Despite the weaker performance of
DNN compared to RF when testing on the simulated data (see Figure
\ref{fig_correlationall}), testing on observational data really show the
power of the algorithm. Nonetheless, all algorithms agree within
$<10\%$. Based on the $f_{1}$ score and \textit{precision}
the methods are all comparable as well. Interestingly, DNN's \textit{ROC
AUC} = 0.589 is the worst among all the methods, just above that of a
classifier with a random guess. 

Judging by the values of the  \textit{precision} which are $\geq 0.95$
for all methods, they satisfy what we require; classifiers that minimise
the number of \HI~poor galaxies incorrectly classified as \HI~rich (\textit{FP}) or in other words with high \textit{precision}.
However, a \textit{specificity} equal to zero implies that all
the negative class instances in the data are incorrectly classified (\textit{FP}), bearing in  mind that only 2$\%$ of this test sample are \HI-poor galaxies. Along with its high \textit{precision}, SVM exhibits the highest \textit{specificity} = 0.714, indicating its robustness, hence the best choice among the algorithms for this test.

We finally note that although $\sim 98\%$ for the RESOLVE galaxies are
\HI-rich, \muf\ sample contains a balanced proportion of $\sim 52\%$
positive class making the training robust against class imbalance effect.  The most important thing is the training part which is achieved using a well balanced sample (50/50 poor-rich), therefore the algorithms are not biased toward any class.   
In Test 4 we will consider a testing set that is more balanced (albeit smaller), which allows us to test our algorithm more fully.

\begin{table}
\caption{Summary of the results when using the simulation trained methods to classify \HI~galaxies in the three different test tests.}
\centering
\begin{tabular}{||c c c c c c||} 
 \hline
  & \textit{Accuracy} & $f_{1}$ & \textit{ROC AUC} & \textit{Precision} & \textit{Specificity}\\ [0.5ex] 
 \hline\hline
\textbf{TEST 1} & & & & & \\
 RF & 0.974 & 0.987 & 0.633 & 0.979 & 0.0 \\
 GRAD & 0.962 & 0.980 & 0.788 & 0.979 & 0.0 \\
 $k$NN & 0.897 & 0.945 & 0.742 & 0.987 & 0.428 \\
 DNN & 0.979 & 0.989 & 0.589 & 0.980 & 0.0\\ 
 SVM & 0.734 & 0.844 & 0.721 & 0.991 & 0.714\\ \hline
\textbf{TEST 2} & & & & & \\
 RF & 0.774 & 0.870 & 0.829 & 0.991 & 0.666 \\
 GRAD & 0.597 & 0.741 & 0.822 & 0.998 & 0.952 \\
 $k$NN & 0.710 & 0.827 & 0.738 & 0.990 & 0.666 \\
 DNN & 0.834 & 0.909 & 0.747 & 0.983 & 0.286\\
 SVM & 0.742 & 0.849 & 0.781 & 0.993 & 0.761\\ \hline
 
\textbf{TEST 3} & & & & & \\
 RF & 0.948 & 0.973 & 0.953 & 1.0 & 1.0 \\ 
 GRAD & 0.881 & 0.937 & 0.970 & 1.0 & 1.0 \\
 $k$NN & 0.882 & 0.937 & 0.900 & 1.0 & 1.0 \\
 DNN & 0.854 & 0.921 & 0.435 & 1.0 & 0.0\\ 
 SVM & 0.848 & 0.917 & 0.893 & 1.0 & 1.0\\ \hline
 
 \textbf{TEST 4} & & & & & \\
 RF & 0.642 & 0.659 & 0.685 & 0.717 & 0.683 \\ 
 GRAD & 0.624 & 0.631 & 0.682 & 0.713 & 0.700 \\
 $k$NN & 0.550 & 0.348 & 0.618 & 0.985 & 0.995 \\
 DNN & 0.732 & 0.761 & 0.666 & 0.767 & 0.418\\ 
 SVM & 0.717 & 0.702 & 0.809 & 0.876 & 0.891 \\ [1ex]
 \hline
\end{tabular}
\label{resolvealfalfa}
\end{table}

\subsection{TEST 2}
\begin{figure}

\includegraphics
{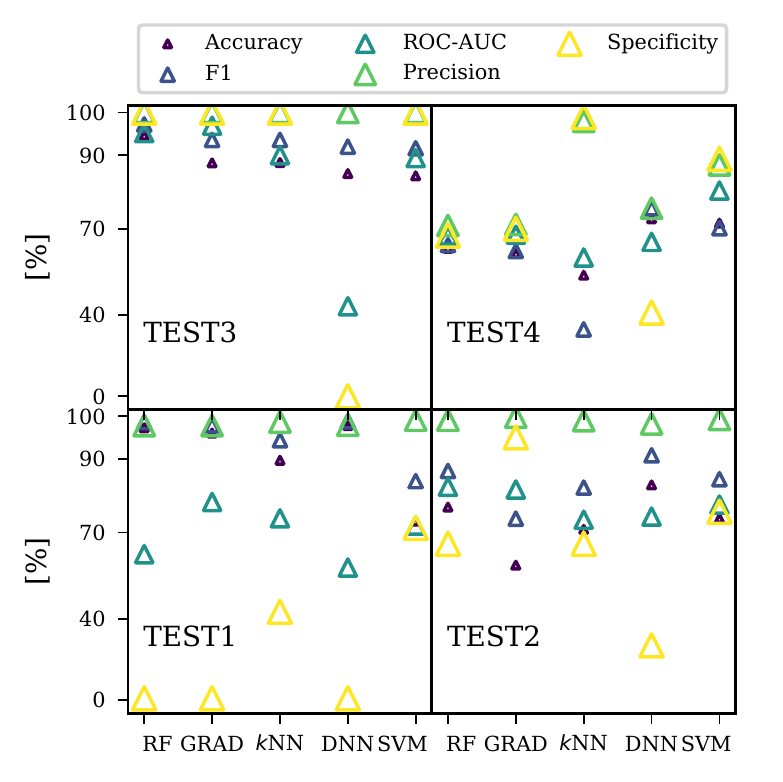}
\caption{Summary of the results when using the simulation trained methods
to classify the \HI\ content of galaxies from three different test
sets from observational data. The y-axes are in exponential scale to prevent
for data point cluttering.}
\label{summaryresults}
\end{figure}

For this second test, we still use the RESOLVE data but consider
\verb'color indices' formed out of SDSS photometric data only,
\textit{i.e.} $\{u,g,r,i,z\}$. In contrast with TEST 1, the results in
Table.~\ref{resolvealfalfa} (Figure \ref{summaryresults}) suggest that,
with the selected inputs features, the methods are capable of better
identifying the gas poor galaxies with \textit{specificity} all above
$0.5$, except for DNN with $0.286$. In terms of {\it Accuracy} and $f_{1}$
scores, DNN is remarkably better and GRAD noticeably worse compared to
RF and $k$NN. Based on ROC AUC, RF and $k$NN score the best and
worst respectively. Based on the value of its \textit{precision} = 0.998, it is tempting to say that GRAD is the best method for this test, however the results suggest that SVM generalises better than GRAD, as indicated by its $f_{1}$ score and \textit{accuracy}. 
It is quite surprising to notice that with the same data (training/test),
decreasing the number of selected features provide better information to
the algorithms such that they get better at classifying the instances
properly \textit{i.e} \textit{precision} (TEST 2) > \textit{precision}
(TEST 1); \textit{specificity} (TEST 2) >  \textit{specificity} (TEST 1).

\subsection{TEST 3}

In this test, we use ALFALFA data and only consider SDSS photometric data
for the input features as in TEST 2. Overall, all the methods perform much
better as suggested by the high values of the metrics considered (see
Table.~\ref{resolvealfalfa}). We note that the training set is the same as
the one used for RESOLVE, hence class imbalance is not an issue that
requires to be alleviated during training. The \textit{precision} and
\textit{specificity} which are both equal to 1 clearly imply that
\textit{FP} is zero, hence class \textbf{0} instances, despite their
relatively low number, are all correctly classified. This applies to all classifiers with the
exception of DNN which has a {\it specificity} = 0.
The results for this test then suggest that our classifiers are capable of
recognizing \HI~rich and \HI~poor galaxies to a very good precision.
The $f_{1}$ scores (all $> 0.9$) of all the learners show that their
\textit{recall}'s are optimised, which also means that \textit{FN}
(\HI~rich galaxies that incorrectly classified as \HI~poor) is minimised.
The relatively higher average precision (\textit{ROC AUC}) of all
classifiers (> 0.9) can indeed be used as an indicator that on average both
\textit{FP} and \textit{FN} are minimised, this is not the case for DNN.
All the trained non-neural network algorithms appear to meet our
requirements but for the sake of comparison, RF method seems to be the best
in this test, with the highest \textit{accuracy} and $f_{1}$ values despite
its \textit{ROC AUC} is only second best. Conversly with the RESOLVE
data, the DNN is definitely not favoured in properly classifying \HI-poor
and \HI-rich galaxies when tests are done with blind survey data such
as ALFALFA. 

\subsection{TEST 4}

We use GASS data \citep{catinella2013galex} for this test, considering SDSS photometric data as input features. Unlike the other samples used for testing so far, all classes (\textbf{0, 1}) are well represented in this dataset, with $56\% $ of this test set are \HI~rich. Although \textit{k}NN exhibits the highest precision and specificity, it does not generalise well, given its relatively low values of both \textit{accuracy} and $f_{1}$. Results suggest that our best classifier for this test is SVM which has a relatively high precision (second best after \textit{k}NN) and its tendency to generalise well as justified by its overall scores. In general, the other classifiers (RF, GRAD and DNN) are also capable of learning the features from the mock data in order to classify the real data, however the neural network model classifies poorly the \HI~poor galaxies (\textit{i.e.} low \textit{specifity} values), as can be noticed in all the tests conducted.

\section{Discussion and Conclusion}\label{conclusiondiscussion}
We have demonstrated in this work that it is possible to classify \HI~galaxies based on
their gas content using both their photometric and environmental data. We have built various
algorithms by training them using large subset of the mock data ($75\%$) from \muf\
simulation. While being sensitive to:
\begin{itemize}
\item the inputs features,
\item type of training (\textit{f-}training or \textit{z-}training),
\item type of class splitting.
\end{itemize} 
the test results, using smaller subset of \muf\ mock data (different from the subset on which
they have been trained), look very promising. For instance, both \textit{Accuracy} and $f_{1}$ score $> 0.9$.

We have shown the good performance of the built classifiers when being tested on real
observation data -- RESOLVE, ALFALFA and GASS surveys -- after training them on the mock data from
\muf\ . Our findings can be summarized as follows:
\begin{itemize}
\item On using \muf\ to both train and test the learners, RF shows the best performance amongst
the learners with an \textit{Accuracy} of $99.00 \%$ \textit{ROC AUC} above $99.96\%$, $f_{1}$
score ~ $99.4\%$ at $z = 1$. Other classifiers like \textit{k-}NN and DNN also perform similarly
well in general, however GRAD method shows poor performance when considering zSMg and fSMg setups.
\item For \textit{z-} training, \textit{Accuracy} and $f_{1}$ score increase from present to
higher redshift. The increase is steeper at $z<0.5$ and flattens out at higher redshift.
This indeed compensates the fact that regressors built in \radpred\ perform best at low redshift
and more poorly with increasing $z$.
\item The performances of the regressors appear to be insensitive to the selected input
features for the training except with the case of GRAD method which struggles to properly
classify the galaxies in the test set when only considering SDSS magnitudes and environmental
information as input features (zSMg and fSMg).
\item The results are affected by the definition of the class of galaxies(BIN, LOW and MULTI).
BIN, which is the type of splitting behind the motivation for this work, corresponds to better results
compared to the other two types of splittings. 
\item Comparing the results corresponding to four different tests using real observational data
from RESOLVE, ALFALFA and GASS surveys, with the exception of DNN as suggested by its low value of \textit{ROC AUC} and zero \textit{specificity}, the classifiers perform best on TEST 3
in which the test set is ALFALFA data and the input features considered are \verb'color indices'
formed out of SDSS magnitudes only. All learners correctly classified the \HI~poor galaxies with
a \textit{specificity} = 1.0 and their precision is also maximised (\textit{precision} = 1.0),
which is what we really aim for. For TEST 3, it is quite clear that most of the errors (if not
all) come from \textit{FN}, {\it i.e.} \HI~rich galaxies misclassified as \HI~poor, although this
quantity is already minimised given the rather high $f_{1}$ score of all the learners.
By comparing TEST 1 and 2, it is clear that using \verb'color indices' from SDSS data only is
the optimal option to better identify the \HI~poor galaxies given the higher \textit{precision}
in TEST 2.
DNN has the highest {\it Accuracy} and $f_{1}$ for TEST 1 and TEST 2, indicative of being
robust in classifying the \HI-rich galaxies. However, DNN fails to achieve a resonable
classification of the \HI-poor galaxies as shown by the low values of {\it Specificity} ($<0.3$)
for all tests. The relatively poor performance of DNN\footnote{Compared to other classifiers in this test.} quantified by the slightly lower values of
{\it Accuracy} and $f_{1}$ for TEST 3 compared to TEST 1 and TEST 2 might be due to the nature
of the test samples. We speculate that the neural network is able to achieve higher performance
in a cleaner set of data such as from the RESOLVE survey but under-perform in a sample from blind
survey data such as ALFALFA. This does not mean the learner itself is not performing well,
it only means that the data to test on are prone to higher systematic errors. 

\item In TEST 4 we use a test sample from GASS, which unlike the other samples used in the first three tests, has a fairly good representation of the two classes (\textit{i.e.}\HI~rich-\HI~poor).  This makes it a good dataset for assessing how well the classifiers are able to apply the learned features from the mock data. Based on the most important performance metric in this study, \textit{k}-NN is the best classifier for TEST 4 with a \textit{precision} = 0.985. It also classifies the \HI~poor galaxies properly as demonstrated by its high \textit{specificity} (0.995). However, even though our purpose is to build a classifier that has a very good \textit{precision} which translates to its ability to correctly classify \HI~rich galaxies, in all kinds of machine learning tasks, the algorithm that can minimise the generalisation errors well is the more preferable. In this case specifically, as the results suggest, SVM proves to be able to generalise well as shown by its \textit{accuracy} (0.717), $f_{1}$ score (0.702), \textit{ROC AUC} (0.809) and both \textit{precision} and \textit{specificity} are the second best.
\item Overall in terms of performance, based on the scores in all tests on real data, we find that SVM is the best classifier as it demonstrates quite well its generalisation ability, learning from simulated data in order to classify real data.
\end{itemize}

With the advent of large \HI~surveys like LADUMA and MIGHTEE, we have presented the possibility
of properly classifying galaxies according to their gas content, using machine learning.
The robustness of our methods lie in the fact that the trained algorithms can learn from mock data
in order to classify galaxies in real surveys, which is indeed a strong asset in the sense that
in reality the lack of enough data to train the methods turns out to be an issue that requires
to be mitigated. Together with the regressors built in \radpred, the classifiers in this work
will form a useful pipeline to create mock \HI\ surveys for assisting with survey design, and eventually, will enable more detailed tests of the input model by comparing observed \HI\ to that predicted from the regressor on a case-by-case basis.

We only analysed the performance of single models in both this work and \radpred. However, the use
of more complex models using ensemble or stacking techniques are increasingly favoured in the
literatures. We will explore such methods in future work despite their level of complexity
as well as their interpretability.
\section*{Acknowledgements}
SA acknowledges financial support from the {\it South African Radio Astronomy Observatory} (SARAO). MR and RD acknowledge support from the South African Research Chairs
Initiative and the South African National Research Foundation.
Support for MR was also provided by the Square Kilometre Array
post-graduate bursary program.
The \muf~simulations were run on the Pumbaa astrophysics
computing cluster hosted at the University of the Western Cape,
which was generously funded by UWC's Office of the Deputy Vice
Chancellor. Additional computing resources are obtained from
the Max Planck Computing \& Data Facility (\url{http://www.mpcdf.mpg.de}) and SARAO.

%%%%%%%%%%%%%%%%%%%%%%%%%%%%%%%%%%%%%%%%%%%%%%%%%%



\bibliographystyle{mnras}
\bibliography{paper_bib} % if your bibtex file is called example.bib

\begin{thebibliography}{}
\makeatletter
\relax
\def\mn@urlcharsother{\let\do\@makeother \do\$\do\&\do\#\do\^\do\_\do\%\do\~}
\def\mn@doi{\begingroup\mn@urlcharsother \@ifnextchar [ {\mn@doi@}
  {\mn@doi@[]}}
\def\mn@doi@[#1]#2{\def\@tempa{#1}\ifx\@tempa\@empty \href
  {http://dx.doi.org/#2} {doi:#2}\else \href {http://dx.doi.org/#2} {#1}\fi
  \endgroup}
\def\mn@eprint#1#2{\mn@eprint@#1:#2::\@nil}
\def\mn@eprint@arXiv#1{\href {http://arxiv.org/abs/#1} {{\tt arXiv:#1}}}
\def\mn@eprint@dblp#1{\href {http://dblp.uni-trier.de/rec/bibtex/#1.xml}
  {dblp:#1}}
\def\mn@eprint@#1:#2:#3:#4\@nil{\def\@tempa {#1}\def\@tempb {#2}\def\@tempc
  {#3}\ifx \@tempc \@empty \let \@tempc \@tempb \let \@tempb \@tempa \fi \ifx
  \@tempb \@empty \def\@tempb {arXiv}\fi \@ifundefined
  {mn@eprint@\@tempb}{\@tempb:\@tempc}{\expandafter \expandafter \csname
  mn@eprint@\@tempb\endcsname \expandafter{\@tempc}}}

\bibitem[\protect\citeauthoryear{Catinella et~al.,}{Catinella
  et~al.}{2010}]{catinella2010galex}
Catinella B.,  et~al., 2010, Monthly Notices of the Royal Astronomical Society,
  403, 683

\bibitem[\protect\citeauthoryear{Catinella et~al.,}{Catinella
  et~al.}{2012}]{catinella2012galex}
Catinella B.,  et~al., 2012, Astronomy \& Astrophysics, 544, A65

\bibitem[\protect\citeauthoryear{{Catinella} et~al.,}{{Catinella}
  et~al.}{2013a}]{Catinella-13}
{Catinella} B.,  et~al., 2013a, \mn@doi [\mnras] {10.1093/mnras/stt1417}, \href
  {http://adsabs.harvard.edu/abs/2013MNRAS.436...34C} {436, 34}

\bibitem[\protect\citeauthoryear{Catinella et~al.,}{Catinella
  et~al.}{2013b}]{catinella2013galex}
Catinella B.,  et~al., 2013b, Monthly Notices of the Royal Astronomical
  Society, 436, 34

\bibitem[\protect\citeauthoryear{Colless, Morganti  \& Couch}{Colless
  et~al.}{1998}]{colless1998looking}
Colless M.,  Morganti R.,   Couch W.,  1998, in eds Morganti R. \& Couch WJ,
  ESO/Australia Workshop, Springer, Pg.

\bibitem[\protect\citeauthoryear{{Dav{\'e}}, {Rafieferantsoa}, {Thompson}  \&
  {Hopkins}}{{Dav{\'e}} et~al.}{2017}]{Dave-17a}
{Dav{\'e}} R.,  {Rafieferantsoa} M.~H.,  {Thompson} R.~J.,   {Hopkins} P.~F.,
  2017, \mn@doi [\mnras] {10.1093/mnras/stx108}, \href
  {http://adsabs.harvard.edu/abs/2017MNRAS.467..115D} {467, 115}

\bibitem[\protect\citeauthoryear{{Davis} et~al.,}{{Davis}
  et~al.}{2011}]{Davis-11}
{Davis} T.~A.,  et~al., 2011, \mn@doi [\mnras]
  {10.1111/j.1365-2966.2011.19355.x}, \href
  {http://adsabs.harvard.edu/abs/2011MNRAS.417..882D} {417, 882}

\bibitem[\protect\citeauthoryear{{Donahue}, {de Messi{\`e}res}, {O'Connell},
  {Voit}, {Hoffer}, {McNamara}  \& {Nulsen}}{{Donahue}
  et~al.}{2011}]{Donahue-11}
{Donahue} M.,  {de Messi{\`e}res} G.~E.,  {O'Connell} R.~W.,  {Voit} G.~M.,
  {Hoffer} A.,  {McNamara} B.~R.,   {Nulsen} P.~E.~J.,  2011, \mn@doi [\apj]
  {10.1088/0004-637X/732/1/40}, \href
  {http://adsabs.harvard.edu/abs/2011ApJ...732...40D} {732, 40}

\bibitem[\protect\citeauthoryear{{Fern{\'a}ndez} et~al.,}{{Fern{\'a}ndez}
  et~al.}{2016}]{Fernandez-16}
{Fern{\'a}ndez} X.,  et~al., 2016, \mn@doi [\apjl]
  {10.3847/2041-8205/824/1/L1}, \href
  {http://adsabs.harvard.edu/abs/2016ApJ...824L...1F} {824, L1}

\bibitem[\protect\citeauthoryear{Folkes et~al.,}{Folkes
  et~al.}{1999}]{folkes19992df}
Folkes S.,  et~al., 1999, Monthly Notices of the Royal Astronomical Society,
  308, 459

\bibitem[\protect\citeauthoryear{{Foltz} et~al.,}{{Foltz}
  et~al.}{2018}]{Foltz-18}
{Foltz} R.,  et~al., 2018, preprint, \href
  {http://adsabs.harvard.edu/abs/2018arXiv180303305F} {} (\mn@eprint {arXiv}
  {1803.03305})

\bibitem[\protect\citeauthoryear{{Fossati} et~al.,}{{Fossati}
  et~al.}{2017}]{Fossati-17}
{Fossati} M.,  et~al., 2017, \mn@doi [\apj] {10.3847/1538-4357/835/2/153},
  \href {http://adsabs.harvard.edu/abs/2017ApJ...835..153F} {835, 153}

\bibitem[\protect\citeauthoryear{Fukugita et~al.,}{Fukugita
  et~al.}{2007}]{fukugita2007catalog}
Fukugita M.,  et~al., 2007, The Astronomical Journal, 134, 579

\bibitem[\protect\citeauthoryear{Gauci, Adami  \& Abela}{Gauci
  et~al.}{2010}]{gauci2010machine}
Gauci A.,  Adami K.~Z.,   Abela J.,  2010, arXiv preprint arXiv:1005.0390

\bibitem[\protect\citeauthoryear{{Haynes} et~al.,}{{Haynes}
  et~al.}{2011}]{Haynes-11}
{Haynes} M.~P.,  et~al., 2011, \mn@doi [\aj] {10.1088/0004-6256/142/5/170},
  \href {http://adsabs.harvard.edu/abs/2011AJ....142..170H} {142, 170}

\bibitem[\protect\citeauthoryear{{Haynes} et~al.,}{{Haynes}
  et~al.}{2018}]{Haynes-18}
{Haynes} M.~P.,  et~al., 2018, \mn@doi [\apj] {10.3847/1538-4357/aac956}, \href
  {http://adsabs.harvard.edu/abs/2018ApJ...861...49H} {861, 49}

\bibitem[\protect\citeauthoryear{{Hess} \& {Wilcots}}{{Hess} \&
  {Wilcots}}{2013}]{Hess-Wilcots-13}
{Hess} K.~M.,  {Wilcots} E.~M.,  2013, \mn@doi [\aj]
  {10.1088/0004-6256/146/5/124}, \href
  {http://adsabs.harvard.edu/abs/2013AJ....146..124H} {146, 124}

\bibitem[\protect\citeauthoryear{Huertas-Company, Rouan, Tasca, Soucail  \&
  Fevre}{Huertas-Company et~al.}{2008}]{HuertasCompany:2007xa}
Huertas-Company M.,  Rouan D.,  Tasca L.,  Soucail G.,   Fevre O.~L.,  2008,
  \mn@doi [Astron. Astrophys.] {10.1051/0004-6361:20078625}, 478, 971

\bibitem[\protect\citeauthoryear{Koekemoer et~al.,}{Koekemoer
  et~al.}{2007}]{koekemoer2007cosmos}
Koekemoer A.~M.,  et~al., 2007, The Astrophysical Journal Supplement Series,
  172, 196

\bibitem[\protect\citeauthoryear{Lahav, Nairn, Sodr{\'e}  \&
  Storrie-Lombardi}{Lahav et~al.}{1996}]{lahav1996neural}
Lahav O.,  Nairn A.,  Sodr{\'e} L.,   Storrie-Lombardi M.,  1996, Monthly
  Notices of the Royal Astronomical Society, 283, 207

\bibitem[\protect\citeauthoryear{Morgan \& Mayall}{Morgan \&
  Mayall}{1957}]{morgan1957spectral}
Morgan W.~W.,  Mayall N.,  1957, Publications of the Astronomical Society of
  the Pacific, 69, 291

\bibitem[\protect\citeauthoryear{Naim, Lahav, Sodre~Jr  \&
  Storrie-Lombardi}{Naim et~al.}{1995}]{naim1995automated}
Naim A.,  Lahav O.,  Sodre~Jr L.,   Storrie-Lombardi M.,  1995, Monthly Notices
  of the Royal Astronomical Society, 275, 567

\bibitem[\protect\citeauthoryear{{Planck} et~al.,}{{Planck} et~al.}{2016}]{-16}
{Planck} et~al., 2016, \mn@doi [\aap] {10.1051/0004-6361/201525830}, \href
  {http://adsabs.harvard.edu/abs/2016A%26A...594A..13P} {594, A13}

\bibitem[\protect\citeauthoryear{{Rafieferantsoa}, {Dav{\'e}},
  {Angl{\'e}s-Alc{\'a}zar}, {Katz}, {Kollmeier}  \&
  {Oppenheimer}}{{Rafieferantsoa} et~al.}{2015}]{Rafieferantsoa-15}
{Rafieferantsoa} M.,  {Dav{\'e}} R.,  {Angl{\'e}s-Alc{\'a}zar} D.,  {Katz} N.,
  {Kollmeier} J.~A.,   {Oppenheimer} B.~D.,  2015, \mn@doi [\mnras]
  {10.1093/mnras/stv1933}, \href
  {http://adsabs.harvard.edu/abs/2015MNRAS.453.3980R} {453, 3980}

\bibitem[\protect\citeauthoryear{Rafieferantsoa, Andrianomena  \&
  Dav{\'e}}{Rafieferantsoa et~al.}{2018}]{rafieferantsoa2018predicting}
Rafieferantsoa M.,  Andrianomena S.,   Dav{\'e} R.,  2018, Monthly Notices of
  the Royal Astronomical Society, 479, 4509

\bibitem[\protect\citeauthoryear{{Rafieferantsoa}, {Dav{\'e}}  \&
  {Naab}}{{Rafieferantsoa} et~al.}{2019}]{Rafieferantsoa-19}
{Rafieferantsoa} M.,  {Dav{\'e}} R.,   {Naab} T.,  2019, \mn@doi [\mnras]
  {10.1093/mnras/stz1199}, \href
  {https://ui.adsabs.harvard.edu/abs/2019MNRAS.486.5184R} {486, 5184}

\bibitem[\protect\citeauthoryear{Slonim, Somerville, Tishby  \& Lahav}{Slonim
  et~al.}{2001}]{slonim2001objective}
Slonim N.,  Somerville R.,  Tishby N.,   Lahav O.,  2001, Monthly Notices of
  the Royal Astronomical Society, 323, 270

\bibitem[\protect\citeauthoryear{{Teimoorinia}, {Ellison}  \&
  {Patton}}{{Teimoorinia} et~al.}{2017}]{Teimoorinia-17}
{Teimoorinia} H.,  {Ellison} S.~L.,   {Patton} D.~R.,  2017, \mn@doi [\mnras]
  {10.1093/mnras/stw2606}, \href
  {http://adsabs.harvard.edu/abs/2017MNRAS.464.3796T} {464, 3796}

\bibitem[\protect\citeauthoryear{York et~al.,}{York
  et~al.}{2000}]{york2000sloan}
York D.~G.,  et~al., 2000, The Astronomical Journal, 120, 1579

\bibitem[\protect\citeauthoryear{Zaritsky, Zabludoff  \& Willick}{Zaritsky
  et~al.}{1995}]{zaritsky1995spectral}
Zaritsky D.,  Zabludoff A.~I.,   Willick J.~A.,  1995, arXiv preprint
  astro-ph/9507044

\makeatother
\end{thebibliography}

%%%%%%%%%%%%%%%%%%%%%%%%%%%%%%%%%%%%%%%%%%%%%%%%%%

% Don't change these lines
%\bsp	% typesetting comment
\label{lastpage}
\end{document}